# Alice Meets Bob: A Comparative Usability Study of Wireless Device Pairing Methods for a "Two-User" Setting

Arun Kumar
Computer Science and Engineering Dept.
Polytechnic Institute of New York University
aashok01@students.poly.edu

Nitesh Saxena
Computer Science and Engineering Dept.
Polytechnic Institute of New York University
nsaxena@poly.edu

Ersin Uzun*
Department of Computer Science
University of California, Irvine
euzun@ics.uci.edu

October 24, 2018

## Abstract

When users want to establish wireless communication between/among their devices, the channel has to be bootstrapped first. To prevent any malicious control of or eavesdropping over the communication, the channel is desired to be authenticated and confidential. The process of setting up a secure communication channel between two previously unassociated devices is referred to as "Secure Device Pairing". When there is no prior security context, e.g., shared secrets, common key servers or public key certificates, device pairing requires user involvement into the process. The idea usually involves leveraging an auxiliary human-perceptible channel to authenticate the data exchanged over the insecure wireless channel.

We observe that the focus of prior research has mostly been limited to pairing scenarios where a single user controls both the devices. In this paper, we consider more general and emerging "two-user" scenarios, where two different users establish pairing between their respective devices. Although a number of pairing methods exists in the literature, only a handful of those are applicable to the two-user setting. We present the first study to identify the methods practical for two-user pairing scenarios, and comparatively evaluate the usability of these methods. Our results identify methods best-suited for users, in terms of efficiency, error-tolerance and of course, usability. Our work sheds light on the applicability and usability of pairing methods for emerging two-user scenarios, a topic largely ignored so far.

## 1 Introduction

Increasing proliferation of personal gadgets (including PDAs, cell-phones, headsets, cameras and media players) equipped with wireless communication (e.g., Wi-Fi, Bluetooth, WUSB) continuously opens up new services and possibilities for ordinary users. There are many usage scenarios where two devices need to "work together." In commonly occurring, so called *"single-user"* scenarios, both communicating devices are controlled by a single user (Alice). Examples include communication between Alice's Bluetooth headset and her cellphone, her PDA and a wireless printer, or her laptop and a wireless access point. On the other hand, *"two-user"* scenarios, whereby two different users (Alice and Bob) control their respective devices, are also fast emerging. Examples include communication between Alice's and Bob's PDAs/laptops/cell phones for sharing files, exchanging digital business cards, multi-player games, messaging, chatting or collaborative applications.

The surge in popularity of wireless devices, however, brings about various security risks. The wireless communication channel is easy to eavesdrop upon and to manipulate, raising the very real threats, notably, of so-called *Man-in-the-Middle* (MiTM) or *Evil Twin* attacks. To mitigate these attacks, secure communication must be first bootstrapped, i.e., devices must be securely paired[1] or initialized.

*Corresponding author.

One of the main challenges in secure device pairing is that, due to sheer diversity of devices and lack of standards, no global security infrastructure exists today and none is likely for the foreseeable future. Consequently, traditional cryptographic means (such as authenticated key exchange protocols) are unsuitable, since unfamiliar devices have no prior security context and no common point of trust. Moreover, the use of a common wireless channel is insufficient to establish a secure context, since such channels are not perceivable by the user.

One valuable and established research direction is the use of auxiliary – also referred to as "out-of-band" (OOB) – channels, which are both perceivable and manageable by the human user(s) who own and operate the devices. An OOB channel takes advantage of human sensory capabilities to authenticate human-imperceptible (and hence subject to MiTM attacks) information exchanged over the wireless channel. OOB channels can be realized using acoustic, visual and tactile senses. Unlike the main (usually wireless) channel, the attacker can not remain undetected if it actively interferes with the OOB channel.

For pairing methods based on OOB channels, some degree of human involvement is essential for achieving security and thus making the user interaction error-free is extremely important. We observe that a large majority of existing device pairing methods (which we review in the following section) are proposed by security professionals without much expertise in usability, leaving the security and performance of the resulting methods unknown at best. Even for the few methods that have been tested for usability, the testing is done in isolation (without facilitating any fair comparison among the methods) or with a limited focus on only single-user scenarios.

**Motivation:** The application domain for secure pairing methods is not just limited to single-user scenarios. Two users often want to exchange files, digital business cards, etc., with each other, when they meet up in person. Main advantage of using Bluetooth or WiFi in such scenarios is that no infrastructure is needed and thus *ad hoc* communication can take place without any extra cost to the users. For this reason, two-user scenarios have been emerging rapidly and are already quite popular, especially in developing countries. Secure pairing of users' devices is a natural and recommended way to prevent any eavesdropping and/or malicious interception during their intended communication.

Many single-user pairing methods have been proposed, each having certain claimed advantages and shortcomings. A single-user pairing method could be directly used in a two-user scenario, only if one of the users operates both the devices. However, it might not always be desirable or feasible to designate both devices to one of the users, e.g., due to security and privacy reasons. In fact, as the survey results of our study show, a majority of users understand such security and privacy implications and would not be willing to share their devices with others, even temporarily during the pairing process. Therefore, it is not clear if existing single-user pairing methods are suitable (and to what extent) when applied in a two-user scenario. In fact, participation of two human users makes the secure pairing process more complicated and potentially error prone.[2] Thus, not all methods might extend well if each device is controlled by a different user.

In short, there is a pressing need to evaluate the applicability and to compare the performance as well as the usability of existing pairing methods for a two-user setting. Such a study is essential to identify pairing method(s) most suitable for everyday users, in terms of efficiency, error-tolerance and of course, usability.

**Our Contributions:** We overview prominent device pairing methods, and identify which of these methods or variants thereof are feasible to use in a two-user setting. We implement the selected methods using a common software platform and conduct a comprehensive and comparative field study, focusing on both usability and security in a two-user setting. Our work helps answer the following question: *what pairing method(s)*

---

[1] We use the term "pairing" to refer to the bootstrapping of secure communication between two devices communicating over a wireless channel

[2] On the other hand, unlike the single-user setting, the devices taking part in the two-user setting are not usually constrained in terms of input/output interfaces. This simplifies the two-user pairing process to a certain extent.

*should be deployed in practice for two-user scenarios?* Without a thorough study, it would be hard to answer this question, simply based on intuition or prior test results on single-user pairing scenarios. Our study yields some interesting results which help us identify most appropriate method(s). Although this paper is less technical in nature than traditional security and applied cryptography research, we believe that the topic is very important and timely since it sheds light on usability in one of the emerging settings where most users (not just specialists) are confronted with security techniques. Also, since most device pairing methods are developed by highly-skilled specialists who are clearly not representative of the general population, there is a certain gap between what seems to be, and what really is, usable. We hope that our work will help narrow this gap.

**Scope:** The scope of this paper is limited only to two-user scenarios. A comprehensive and comparative usability evaluation of pairing methods for single-user scenarios is of independent interest, which has been addressed in various recent work [12, 10, 8].

**Paper Organization:** The rest of the paper is organized as follows: Section 2 reviews notable cryptographic protocols and relevant device pairing methods. Next, Section 3 discusses some pre-study design choices and criteria, followed by usability testing details in Section 4. Next, we discuss, in Section 5 our interpretations of the results obtained in the course of the study and conclude with the summary and future work in Section 6.

## 2 Background

In this section, we describe notable relevant cryptographic protocols and device pairing methods. The term *cryptographic protocol* denotes the entire interaction involved, and information exchanged, in the course of device pairing. The term *pairing method* refers to the pairing process as viewed by the user, i.e., the sequence of user actions. As discussed later on, a single cryptographic protocol can be coupled with many pairing methods.

### 2.1 Cryptographic Protocols

One simple protocol was suggested in [1], where devices A and B exchange their respective public keys $pk_A$, $pk_B$ over the insecure channel and the corresponding hashes $H(pk_A)$ and $H(pk_B)$ – over the OOB channel. Although non-interactive, the protocol requires $H()$ to be a (weakly) collision-resistant hash function and thus needs at least 80 bits of OOB data in each direction. MANA protocols [6] reduce the size of OOB messages to $k$ bits while limiting attacker's success probability to $2^{-k}$. However, these protocols require a stronger assumption on the OOB channel: the adversary is assumed to be incapable of delaying or replaying any OOB messages.

In [29], the author presented the first protocol based on Short Authenticated Strings (SAS), which limits attack probability to $2^{-k}$ for a $k$-bit OOB channels, even when the adversary can delay/replay OOB messages. This protocol utilizes commitment schemes (which can be based upon hash functions such as SHA-1, MD5) and requires 4-round of communication over the wireless channel. Subsequent work ([14] and [21]) developed 3-round SAS protocols. Both are modifications of [29] and both use (although in different way) a universal hash function in the computation of SAS messages. Recently, authors of [20] proposed a more efficient SAS protocol requiring a $k$-bit SAS message in one direction, but only a 1-bit SAS message in the other. In this protocol, device A sends its $k$-bit SAS message to device B; B checks whether the received copy matches its own, and communicates the result to A with a 1-bit SAS message. As discussed later, this protocol is utilized in a number of pairing methods we tested.

### 2.2 Device Pairing Methods

Over the recent year, a number of pairing methods have been proposed. They operate over different OOB channels, use different cryptographic protocols and offer varying degrees of usability. Recall that these methods have so far been proposed in the context of a single-user setting, and not the two-user setting, which is the focus of this paper. Later in Section 3, we will discuss the applicability of these methods to the two-user setting.

The initial attempt to address the device pairing problem in the presence of MiTM attacks was "Resur-recting Duckling" [26]. It requires standardized physical interfaces and cables. Although it was appropriate in the 1990-s, this is clearly obsolete today, due to the greatly increased diversity (and decreased size) of devices and the requirement of a physical equipment (i.e., a cable) which defeats the purpose and convenience of wireless connections.

"Talking to Strangers" [1] was another early method, which relies on infrared (IR) communication as the OOB channel and requires almost no user involvement, except for initial setup. Moreover, it has been experimented with user (unlike many other methods), as reported in [2]. However, this method is deceptively simple since IR is line-of-sight and, setting it up requires the user to find IR ports on both devices – not a trivial task for many users – and align them. Also, despite its line-of-sight property, IR is not completely immune to MiTM attacks. Another drawback is that IR has been largely displaced by other wireless technologies (e.g., Bluetooth) and is available on few modern devices.

Another early approach involves image comparison. It encodes the OOB data into images and asks the user to compare them on two devices. Prominent examples include "Snowflake" [7], "Random Arts Visual Hash" [22] and "Colorful Flag" [4]. Such methods, however, require both devices to have displays with sufficiently high resolution. Applicability is therefore limited to high-end devices, such as: laptops, PDAs and certain cell phones. These methods are based on the protocol proposed in [1] which was reviewed earlier. A more practical approach, based on SAS protocols [21, 14], suitable for simpler displays and LEDs has been investigated in [28].

More recent work [19] proposed the "Seeing-is-Believing" (SiB) pairing method. In its simplest instantiation, SiB requires a uni-directional visual OOB channel for one-way authentication: one device encodes OOB data into a two-dimensional barcode which it displays on its screen and the other device "reads it" using a photo camera, operated by the user. At a minimum, SiB requires one device to have a camera and the other – a display for uni-directional authentication and both devices to have a camera and display for bi-directional authentication. Thus, it is not suitable for small or low-end devices.[3] From the user's perspective, SiB is a relatively undemanding pairing method as user actions amount to taking a photo of a barcode.

A related approach, called "Blinking Lights" has been explored in [20]. Like SiB, it uses the visual OOB channel and requires one device to have a continuous visual receiver, e.g., a light detector or a video camera. The other device must have at least one LED. The LED-equipped device transmits OOB data via blinking while the other receives it by recording the transmission and extracting information based on inter-blink gaps. As in SiB, the receiver device indicates success/failure to the user who, in turn, informs the other to accept or abort.

Quite recently, [23] developed a pairing method based on synchronized audio-visual patterns. Three proposed methods, "Blink-Blink", "Beep-Beep" and "Beep-Blink", involve users comparing very simple audiovisual patterns, e.g., in the form of "beeping" and "blinking", transmitted as simultaneous streams, forming two synchronized channels. One advantage of these methods is that they require devices to only have two LEDs (one of which is to ensure synchronization) or a basic speaker.

Another recent method is "Loud-and-Clear" (L&C) [16]. It uses the audio (acoustic) OOB channel along with vocalized MadLib sentences or phrases which represent the digest of information exchanged over the main wireless channel. There are two L&C variants: "Phrase-DS" and "Phrase-SS". In the latter the user compares two vocalized sentences and in the former – displayed sentence with its vocalized counterpart. Minimal device requirements include a speaker (or audio-out port) on one device and a speaker or a display on the other. The user is required to compare the two respective (vocalized and/or displayed) MadLib sentences and either accept or abort the pairing based on the outcome of the comparison. As described in [16], L&C is based on the protocol of [1]. In this paper, to reduce the number of words in the MadLib sentences, we use the L&C variant based on SAS protocols [21, 14]. The third variant of L&C, "Phrase-DD," simply involves displaying the sentences on two devices, which the user is asked to compare.

Some follow-on work (HAPADEP [25]) considered pairing devices that – at least at pairing time – have no common wireless channel. HAPADEP uses pure audio to transmit cryptographic protocol messages and requires the user to merely monitor device interaction for any extraneous sounds or interference. It requires both devices to have speakers and microphones. To appeal to more common setting (one where a common wireless channel is available), we employ a HAPADEP variant, we call "Over-Audio." This variant uses the wireless channel for cryptographic protocol messages and the audio – as the OOB channel. In it, only one device needs a speaker and the other – a microphone. Also, the user is not involved in any comparisons.

---

[3]Albeit, the display requirement can be relaxed in case of a printer; alternatively, a camera-equipped device can snap a photo of a barcode sticker affixed to the *dumber* device.

An experimental investigation [27] presented the results of a comparative usability study of simple pairing methods for devices with displays capable of showing a few (4-8) decimal digits of OOB data. In the "Compare-Confirm" or "Numeric-Compare" approach, the user simply compares two 4-, 6- or 8-digit numbers displayed by devices. In the "Select-Confirm" approach, one device displays to the user a set of (4-, 6- or 8-digit) numbers, the user selects the one that matches a single such number displayed by the other device. In the "Copy-Confirm" approach, the user copies a number from one device to the other. The last variant is "Choose-Enter" which asks the user to pick a "random" 4-to-8-digit number and enter it into both devices. All of these methods are undoubtedly simple, however, as [27] indicates, Select-Confirm and Copy-Confirm are slow and error-prone. Furthermore, "Choose-Enter" is insecure since studies show that the quality of numbers (in terms of randomness) picked by the average user is very low.

Yet another approach: "Button-Enabled Device Authentication" [24] suggests pairing devices with the help of user button presses, thus utilizing the tactile OOB channel. This method has several variants: "BEDA-Blink", "BEDA-Beep", "BEDA-Vibrate" and "BEDA-Buttons". In the first three variants, respectively, based on the SAS protocol variant [20], the sending device blinks its LED (or beeps or vibrates) and the user synchronously presses a button on the receiving device. Each 3-bit block of the SAS string is encoded as the delay between consecutive blinks (or beeps or vibrations). As the sending device blinks (or beeps or vibrates), the user presses the button on the other device thereby transmitting the SAS from one device to another. In the BEDA-Buttons variant, which can work with any PAKE protocol (e.g., [3]) the user simultaneously presses buttons on both devices and random user-controlled inter-button pressing delays are used as a means of establishing a common secret.

A very different OOB channel was considered in "Smart-Its-Friends" [13]: a common movement pattern is used to communicate a shared secret to both devices as they are shaken together by the user. A similar approach is taken in "Shake Well Before Use" [17]. Both techniques require devices to be equipped with 2-axis accelerometers. Although some recent mobile phones (e.g., iPhone) are equipped with it, accelerometers are not common on other devices.

There are also other methods involving technologies that are more relatively expensive and uncommon. To summarize a few. [9] suggested using ultrasound as the OOB channel. A related technique uses laser as the OOB and requires each device to have a laser transceiver [18]. However, the hardware needed for these methods are not readily available in many current devices and are not expected to be deployed soon.

## 3 Study Preliminaries

This section discusses selection criteria for tested methods and devices as well as the architecture of our software platform.

### 3.1 Methods to be Tested

As follows from our overview in the previous section, there is a large body of prior research on secure device pairing, comprised of a wide range of methods. As mentioned earlier, all of these methods were proposed in the context of a single-user pairing setting. Some of these methods have been evaluated in terms of their usability. These include Talking-to-Strangers (Network-in-a-Box) [2], Compare-Confirm, Copy-Confirm and Choose-Enter [27], as well as all four variants of BEDA [24] and Blink-Blink, Beep-Beep and Beep-Blink combinations [23].

There are more than twenty methods, counting variations, in literature. However, some of these methods have very limited use cases due to either requiring both devices to be controlled by the same user during the pairing (e.g., accelerometer based methods such as [17]) or requiring hardware that is not ubiquitous among wireless enabled devices. Some methods also have stronger assumptions about the OOB channel and require it to be confidential (e.g., BEDA-Buttons variant of [24]). Notice that secret OOB channels are hard to achieve in real-life since a close-by attacker can eas-ily eavesdrop on any human perceptible channel (e.g., by shoulder surfing the pairing process).

We believe that it is very difficult to test all available methods in one single study and hope that the results will yield meaningful comparative usability metrics. Obstacles like varying security assumptions about the OOB channel among different methods and possible user fatigue from including too many methods would undermine the results of such study. Consequently, we have to cull the number of methods down to a more manageable number, eliminating those that are obsolete, deprecated based on prior evaluations or unrealistic on their OOB assumptions. Of course, we also eliminated any methods that are limited to a single-user setting only. The following methods are excluded from our study:

- Resurrecting-Duckling[26]: obsolete due to physical equipment, i.e., cable, requirement.

- Talking-to-Strangers[1]: obsolete since IR ports are not secure against MiTM attacks and they have become uncommon.

- Choose-and-Enter[27], BEDA-Buttons [24]: requires a secret OOB channel and performed poorly in prior evaluations.

- Beep-Beep[23]: performed poorly in prior evaluations due to user annoyance and high error rate.

- Blink-Blink[23], Image Comparison[7, 22, 4]: do not extend well to a two-user setting, as the two devices need to be placed adjacent to each other or temporarily exchanged between users; also, varying resolutions on devices makes image comparison a burdensome and error-prone task.

- Seeing-is-Believing[19], Blinking Lights[20]: require photo or video cameras on devices and do not extend well to the two-user setting due to the need for close proximity between the devices; also cameras are not ubiquitous interfaces except for mobile phones.

- BEDA-Vibrate[24]: vibration is not a common interface, except for mobile phones; also it is hard for one user to sense the vibration on another user's device, making this method unusable in a two-user setting.

- Smart-its-Friends[13], Shake-Well-Before-Use[17]: requires one user to hold and control both devices and thus do not extend to two-user settings.

- Ultrasound[9] and laser[18] based methods: requires interface and hardware capabilities that are not common across devices.

Remaining methods have been included in our two-user study. However, we had to slightly modify or update certain methods to standardize the OOB assumptions and security among the tested methods. In particular, we have updated all methods to be based on a SAS protocol for better efficiency and unified security assumptions. This update resulted in a slightly changed user interaction in BEDA-Beep, BEDA-Blink and Over-Audio methods from their original proposals. A brief description of user interactions involved when Alice is pairing her device A with Bob's device B in the methods we tested is as follows.

- **BEDA-Beep**: As and when A beeps, Bob presses a button on B synchronously; B indicates the result of pairing on its screen; Bob indicates the same result to Alice; Alice accepts or rejects pairing on A accordingly. The time intervals between successive beeps encodes the 15-bit SAS value.

- **BEDA-Blink**: As and when A blinks/flashes its screen, Bob presses a button on B synchronously; B indicates the result of pairing on its screen; Bob indicates the same result to Alice; Alice accepts or rejects pairing on A accordingly. The time intervals between successive blinks encodes the 15-bit SAS value.

- **Beep-Blink**: While A is blinking/flashing its screen and B is beeping, Alice compares the blinking/flashing of A with the beeping of B and Bob compares the beeping of B with the blinking/flashing of A; based on the comparison, both Alice and Bob accept or reject the pairing on A and B, respectively. The on-off blinking/beeping encodes a 15-bit string.

- **Over-Audio**: A plays an audio that has 15-bit string encoded into it and B records this audio; B indicates the result of pairing on its screen; Bob indicates the same result to Alice; Alice accepts or rejects pairing on A accordingly.

- **Numeric-Compare**: A and B display a 5-digit number (each) on their respective screens; Alice compares the number displayed on A with the number displayed on B (with Bob's help); Bob compares the number displayed on B with the number displayed on A (with Alice's help); based on the comparison, both Alice and Bob accept or reject the pairing on A and B, respectively.

- **Phrase-DD**: A and B display 3-word phrases on their respective screens; Alice compares the phrase displayed on A with the phrase displayed on B (with the help of Bob); Bob compares the phrase displayed on B with the phrase displayed on A (with the help of Alice); based on the comparison, both Alice and Bob accept or reject the pairing on A and B, respectively.

- **Phrase-DS**: A displays a 3-word phrase on its screen and B "speaks-out" a 3-word phrase; Alice compares the phrase displayed on A with the phrase spoken by B; Bob compares the phrase spoken by B with the phrase displayed on A (with Alice's help); based on the comparison, both Alice and Bob accept or reject the pairing on A and B, respectively.

- **Phrase-SS**: A speaks out a 3-word phrase and then B speaks out a 3-word phrase; Alice compares the phrase spoken by A with the phrase spoken by B; Bob compares the phrase spoken by B with the phrase spoken A; based on the comparison, both Alice and Bob accept or reject the pairing on A and B, respectively.

- **Copy-Confirm**: A displays a 5-digit number on its screen; Alice indicates the number to Bob and he inputs it on B; B indicates the result of pairing on its screen; Bob indicates the same result to Alice; Alice accepts or rejects pairing on A accordingly.

### 3.2 Test Devices to be Used

In a single-user setting, one of the devices might be constrained in terms of input and output interfaces. For example, a headset, while being paired with a cell phone and an access point, while being paired with a laptop, are constrained devices (with no display, keypad). On the other hand, both devices participating in a two-user communication setting are "personal" devices (such as PDAs, cell phones, laptops) and therefore would not be constrained. These devices are generally equipped with at least a full display and a keypad.

We wanted to simulate, as closely as possible, common two-user pairing scenarios. To this end, for our entire study, we used two Nokia cellphones models:[4] N73 and E61, as test devices. Both models have been released two years ago (in 2006) and hence do not represent the cutting edge technology. This was done on purpose in order to avoid devices with expensive features as well as processors faster than those commonly available at present.

Another reason for choosing these devices is the plethora of *common* interfaces available on them. Recall that our goal is to test many methods utilizing many different OOB channels, including: audio and visual and tactile. For each of these channels, some methods need user-input, user-output or both. The audio channel can require: speaker, beeper or a microphone. The visual channel requires an LED or a screen, whereas, the tactile channel can require: button or keypad. Our test devices have all these features which allows testing all methods consistently. (Otherwise, changing devices across methods would seriously undermine the credibility of results.) Specifically, both N73 and E61 have the following features relevant to our tests:

- Input Interface: keypad (subsumes button), microphone

- Output Interface: speaker (subsumes beeper), color screen (subsumes LED)

- Wireless: Bluetooth

In all tests, Bluetooth was used as the wireless (human-imperceptible) channel. We consider this choice to be natural since Bluetooth is widely available and inexpensive.

For methods that involve beeping, the general-purpose speaker is trivial to use as a beeper. Whenever a button is needed, one of the keypad keys is easily configured for that purpose. A picture of a bright LED showed on the screen in place of a blinking LED.

---
[4]For N73 specs, see: www.nokiausa.com/A4409012, and for E61 – europe.nokia.com/A4142101.

Utilizing the whole screen rather than an LED was a natural and an obvious choice for two-user pairing.

## 3.3 Implementation Details

In comparative usability studies, meaningful and fair results can only be achieved if all methods are tested under similar conditions and settings. In our case, the fair comparison basis is formed by: (1) keeping the same test devices, (2) employing consistent GUI design practices (e.g., safe defaults), and (3) unifying targeted (theoretical) security level for all methods. Our goal is to isolate – to the extent possible – user interaction in different methods as the only independent variable throughout all tests. Minimizing any experimenter-introduced, systematic and cognitive bias is also important. In particular, we randomize test order, avoid close physical proximity and interaction between the participants and the experimenter, and automate timing and logging to minimize errors and biases.

Some of the tested methods already had prior working prototype implementations. However, these were mostly developed by their authors who aimed to demonstrate implementation feasibility. Consequently, such prototypes are often: incomplete, buggy and/or fragile as well as very dependent on specific hardware/software platforms. It is nearly impossible to provide a uniform testing environment using available prototypes. Modifying them or implementing each from scratch is also not an option, due to the level of effort required. For stand-alone applications, implementing only the user interface is usually enough for the purposes of usability testing. However, distributed applications, such as secure device pairing, need more than just user interface, since a realistic user experience is unattainable without any connection between devices.

To achieve a unified software platform, our implementation used the open-source comparative usability testing framework developed by Kostiainen, et al. [11]. It provides basic communication primitives between devices as well as automated logging and timing functionality. However, we still had to implement separate user interfaces and simulated functionality for all tested methods. We used JAVA-MIDP to implement all methods and created several test-cases for "no-attack" and "under-attack" scenarios. (The term *attack* is limited to MiTM/Evil-Twin attacks in this context).

For all methods, we kept the SAS string length constant at 15 bits. We also tried to keep all user interfaces similar, while applying same design practices, i.e, safe-default selection prompts, clear instructions, simple language, etc. All methods are precisely timed from the start to the end of user interaction.

We believe that, in our implementation, user experience and interaction model are very realistic. For most methods tested, the only difference between our variant and a real method is that we omitted the initial rounds of the underlying cryptographic protocol (e.g., SAS) that use the wireless channel, i.e., do not involve the user. Instead, our implementation supplies devices with synthetic SAS strings to easily simulate normal and MiTM attack scenarios. However, since messages over the wireless channel are completely transparent to the user, our *simulation*, from the user's perspective very closely resembles the real-life version.

## 4 Usability Testing Details

Having implemented all selected pairing methods on a common platform, we are ready to start the usability study for the two-user setting. Our goal is to evaluate and compare pairing methods we identified (and listed in Section 3.1) with respect to the following factors:

1. Efficiency: time it takes to complete each method

2. Robustness: how often each method leads to false positives (or rejection of a successful pairing instance) and false negatives (or acceptance of an unsuccessful pairing instance, i.e., pairing instance that is not between the intended devices). Following the terminology introduced in [27], we will refer to the errors in the former category as *safe errors* and the latter as *fatal errors*.

3. Usability: how each method fares in terms of user burden (i.e., ease-of-use perception) and personal preference.

4. User Interactions: how the two users interact in order to perform the various steps involved in each method, and in particular, would they hand-in their devices to each other.

### 4.1 Study Participants

For our tests, we recruited 40 participants[5] and clearly at a given point of time two participants had to be present to complete the tests. The study lasted over a period of more than two months. Each pair of participants was chosen very carefully as we required them to have varying trust relationships with each other, ranging from being strangers, to acquaintances to close buddies. Each pair of users was briefed on the estimated amount of time required to complete the tests. The participants chosen were mostly university students both at the undergraduate and graduate level. Thus our study represents only the first step towards identifying methods suitable for the broad cross-section of user population.

| | | |
|---|---|---|
| **Age** | 18-24 | 95% |
| | 25-29 | 5% |
| **Gender** | Female | 35% |
| | Male | 65% |
| **Avg. Computer Experience** | 8.5 years (2.14 StDev) | |
| **Avg. Computer Usage Per Day** | 7.75 hours (3.82 StDev) | |

Figure 1: User Demographics

We prepared two questionnaires: *background* – to obtain user demographics and *post-test* – for user feedback on methods tested. The user demographics are as shown in Figure 1.

None of the study participants reported any physical impairments that could interfere with their ability to complete given tasks. The gender split was: 65% male and 35% female as shown in the figure. Gender and other information was collected through *background questionnaires* completed prior to testing. Also collected prior to testing was the information on whether the participants knew each other and if yes, how well did he/she know the other person.

**Trust relations between participants:** Among the 20 subject pairs, 5 of them have not met before (and thus were complete strangers), 5 of them were close friends, and the remaining 10 were friends or colleagues but they did not consider each other as very close friends. In order to gain some insight into the trust relations and the acceptable interaction between the subject pairs, we asked them whether they would consider temporarily handing their device to the other person in order to initiate a secure connection that they can later use to exchange files, messages or play games. We also asked their reasoning and concerns related to their answers.

Not surprisingly, all the 5 pairs that have not met before said they wouldn't consider any physical exchange of the devices as an acceptable interaction. The two main concerns we identified in those pairs was the security of the device and the data it stores as well the unpleasant social situation it may create. On the other hand, 4 out of 5 pairs that are close friends and have known each other well did not state any privacy concerns and said they would physically exchange their devices during the pairing if needed. Among the 10 pairs that are friends or colleagues, 6 of them expressed serious concerns about any physical device exchange and considered it unacceptable and their reasonings were similar to the first group.

From the observed trust relations and concerns expressed by our subjects, we can conclude that any method that needs physical exchange of devices is unacceptable in many scenarios where the owners of the devices *do not know each other very well*. Moreover, it may still be problematic in some situations even if the device owners knew each other. Among the 8 pairs who were not reluctant to exchanging devices, the relationship between the users played a strong role. Surprisingly, 5 among the 8 pairs only considered friends as the acceptable social group to temporarily exchange devices and even excluded family members. The remaining 3 considered both family and the friends as acceptable. However, we believe that the observed strong tendency to share devices with friends rather than with the family members was perhaps due to the biased sampling of our subjects, i.e., mostly unmarried college students.

### 4.2 Test Cases

There were a total of 9 tests, which each of our 20 subject pairs performed. The test cases were designed in such a manner that the attacks will show up proba-bilistically, meaning that the user will not see both no-attack and under-attack scenario for each method but they will see either of them with half a chance. This way we prevent the users from expecting one no-attack and one under-attack test case for each method. This also reduces the number of tests to be performed by the users.

---
[5]This was one of the challenges in pursuing a two-user study – twice as many participants were needed compared to the single-user testing. It is well-known that a usability study performed by 20 set of participants captures over 98% of usability related problems [5]

### 4.3 Testing Process

Our study was conducted in a variety of campus venues including, but not limited to: student laboratories, cafés, student dorms/apartments, classrooms, office spaces and outdoor terraces. This was possible since the test devices were mobile, test set-up was more-or-less automated and only a minimal involvement from the test administrator was required.

After giving a brief overview of our study goals, we asked the participants to fill out the background questionnaire in order to collect demographic information. Both the participants in each test jointly filled up the questionnaire. In this questionnaire, we also asked the participants whether they suffer(ed) from any visual or hearing impairments, or have any condition that may interfere with their operating of devices or their reflexes. After a short interview about the relationship between each pair of subjects, they were given a brief introduction to the cell-phone devices used in the tests.

Each pair of users was given two devices (one per user) and asked to follow the on-screen instructions shown before each task to complete it. The two users were to mutually perform the tests. A close watch was made to observe whether the two users exchanged their respective devices to carry out the tests.

User interactions throughout the tests were observed by the test administrator and timings were logged automatically by the testing framework. After completing the tasks, each pair jointly filled out the post-test questionnaire form, in which they provided their feedback on the various methods tested and also if they found any particular test to be difficult. The pairs were also given a few minutes of free discussion time, where they discussed with the test administrator their experience with the various methods tested.

### 4.4 Test Results

We collected data in two ways: (1) by timing, observing and logging user interaction, and (2) via question-naires and free-form interviews.

| Method | Time (seconds) | Fatal Error rate | Safe Error rate | Avg # of trials until success |
|---|---|---|---|---|
| BEDA-Beep | 40.43 | 0.00 | 0.14 | 1.14 |
| BEDA-Blink | 96.00 | 0.00 | 0.10 | 2.20 |
| Beep-Blink | 45.11 | 0.09 | 0.11 | 1.20 |
| Over-Audio | 18.75 | 0.00 | 0.00 | 1.13 |
| Numeric-Compare | 12.50 | 0.00 | 0.10 | N/A |
| Phrase-DD | 11.44 | 0.00 | 0.00 | N/A |
| Phrase-DS | 21.45 | 0.00 | 0.00 | N/A |
| Phrase-SS | 38.71 | 0.00 | 0.00 | N/A |
| Copy-Confirm | 17.00 | 0.17 | 0.00 | N/A |

Figure 2: Summary of Logged Data

For each method, completion times, errors, actions and the playcount, i.e., the number of trials it took before successful pairing was established, were automatically logged by the software. The logged data is summarized in Figure 2. The average timing information (with standard deviations) is graphed in Figure 3. Also graphed, in Figure 4, is the playcount for the applicable methods.

We also observed the user interactions while each pair of users was performing the various steps involved in each tested method. In general, we observed that the subjects often decided the outcome of pairing based on mutual agreement, which, we believe, may have helped to reduce errors in most of our comparison-based methods. We did not observe any statistical correlation between the closeness of the relationship between the participants and their interactions during the pairing process. Observed interactions for each method are summarized below. (Assume Alice is pairing her device A with Bob's device B)

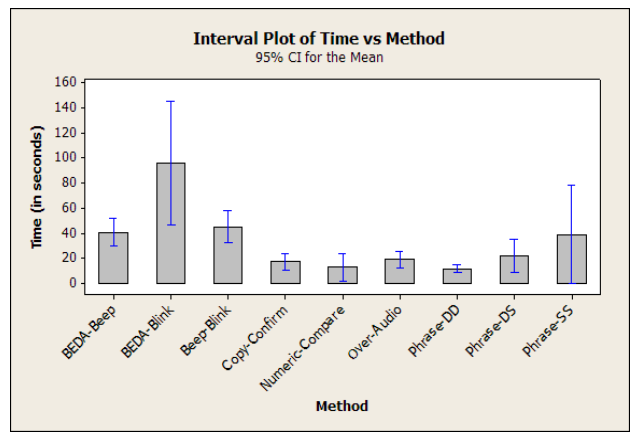

Figure 3: Time-to-Completion for Successful Pairing

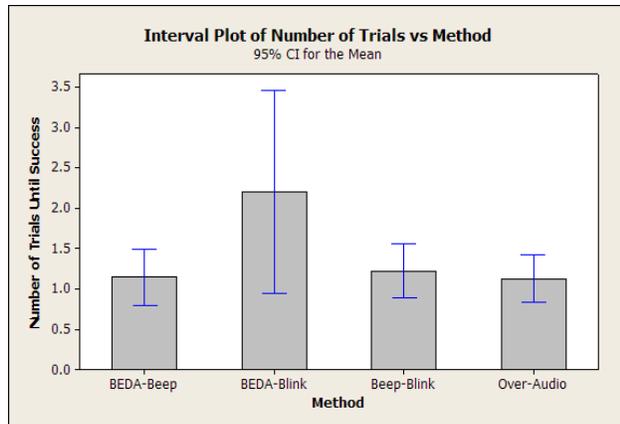

Figure 4: Interval Plot of Playcount

- BEDA-Beep: The user responsible for pressing the button (Bob) would listen carefully for the beeping on the other device (A) and synchronously press any button on B. The user of the beeping device (Alice) moved closer to Bob within a distance of about 1-2 feet so that the beeping sound could be heard clearly by him. Alice was also noticing if Bob was synchronizing the beeping with the button press. Once finished with this phase, Bob verbally notified Alice to accept or reject the pairing, based on the result shown on B.

- BEDA-Blink: The user of the blinking device (Alice) would show her device to the other user (Bob) and Bob would press the button in synchronization with the blinking. The users were again within a distance of 1-2 feet. Once finished with this phase, Bob verbally notified Alice to accept or reject the pairing, based on the result shown on B.

- Beep-Blink: After starting the pairing process, both Alice and Bob compared the blinking/beeping on their own device with the beeping/blinking on the other device. This required the two users to be within touching distance of each other so that both could watch as well as listen to the flashing screen and beeping on the two devices, respectively. At the end, both users will accept or reject the pairing on their respective devices, based on their mutual judgement of whether blinking/beeping was synchronized or not.

- Over-Audio: In this method, the role of the two users was quite "passive." Alice's device would start to play an audio encoding a bitstring and Bob's device would automatically record it. After the audio transfer was over, Bob would (verbally) tell Alice to accept or reject depending on what his screen indicated.

- Numeric-Compare: In this method, both Alice and Bob either spell out or show the number displayed on the screens of their devices, compare the two and mutually accept or reject the pairing on their respective devices.

- Phrase-DD: Similar to Numeric-Compare, both Alice and Bob either spell out or show the sentence displayed on the screens of their devices, compare the two and accept or reject the pairing on their respective devices, based on mutual agreement.

- Phrase-DS: This method involves the user (Alice) to listen carefully to the sentence spelled out by the device of other user (Bob) and then compare it with the sentence displayed on the screen of her device, and vice versa. For this to take place, Bob would take his device closer (about 1-2 feet) to Alice's device, in order for her to be able to listen clearly to the spoken sentence. Alice and Bob mutually accept or reject the pairing on their respective devices, following a short discussion.

- Phrase-SS: This method involves both the devices to speak out a sentence each. Thus, when the devices speak out, both the users would lean towards the devices to listen to them carefully. After listening to both the sentences, they would analyze if the sentences were the same. They would then accept or reject the pairing on their respective devices, after verbally confirming with each other.

- Copy-Confirm: Alice would either read out the number displayed on A or directly show the screen displaying the number to Bob. After feeding in the number on his device, Bob would verbally notify Alice to accept or reject, depending on what his screen directs him to do.

Finally, through the post-test questionnaire, we solicited user opinions about all tested methods. Each pair of users rated each method on a 6 level Likert scale[15] over various statements such as the method is easy to use, professional, fun to use, tiring, takes too long to complete, and error prone. The ease-of-use and other attribute ratings are graphed in Figure 5.

## 5 Interpreting Results

In this section we attempt to interpret the results of our study. We first consider various mechanical data, i.e., time to completion and error rates. We then analyze the perceived qualitative aspects of the methods based on collected user ratings.

### 5.1 Interpreting Time and Error Results

Our results reflected in Figure 2 and Figure 3, prompt a number of observations.

**Completion time:** The logical way to interpret the completion time is by looking at it under normal circumstances, i.e., when no active or passive attacks occur, as this is what users will normally experience. Thus, we only considered the no-attack test cases while calculating the average completion time. Based on this performance metrics, tested methods fall into three speed categories: fast (less than 20 secs), moderate (between 20 and 30 secs) and slow (more than 30 secs). The fastest method is Phrase-DD at 11.44 secs for a successful outcome and it is closely followed by the Numeric-Compare at 12.5 secs. Copy-Confirm and Over-Audio methods are next taking 17 and 18.75 secs, respectively. Phrase-DS comes in at 21.45 and its performance is considered moderate and acceptable. The slow category includes the rest, ranging from Phrase-SS (38.71 secs) to BEDA-Blink which takes a whooping 96 secs.

**Error Rates:** As explained in section 4, *fatal errors* have a grave effect on security as they would result in a MiTM attack to be successful. On the other hand, *safe errors* do not constitute an immediate security threat but will create user annoyance as the pairing process has to be repeated, thus prolonging the completion time. However, in addition to poor usability experience, safe errors may eventually threaten security as too much user annoyance may result in careless behavior that can yield to fatal errors.

In our tests, most methods, except Copy-Confirm and Beep-Blink, fare well reporting no fatal errors. Copy-Confirm and Beep-Blink suffered from fatal error rates of 17% and 9% respectively, which constitutes a significant vulnerability in the context of security applications. BEDA-Beep, BEDA-Blink, Beep-Blink and Numeric-Compare methods all yield more than 10% error rates but for safe errors which are considered to be not directly threatening the security. However, as discussed earlier, safe errors are indication of poor usability and may eventually have adverse effects on security as well as efficiency due to the user annoyance (e.g., after many trials with unsuccessful outcome, an annoyed user may just accept the connection without checking or despite the non-matching SAS values).

Fortunately, it turns out that the fastest method is also one of the error-free methods. Taking both speed and error-rate into account, the overall best method is clearly Phrase-DD, followed by Over-Audio and Phrase-DS. Although Numeric-Compare is also quite fast, there is little motivation for using it over Phrase-DD. The reason is simple: both methods require the same hardware on devices (basic displays) and Phrase-DD provides lower error rates and takes about the same time as Numeric-Compare (this also confirms our intuition that users are better at interpreting phrases over numbers). Thus, for two-user pairing scenarios, where both devices are equipped with decent quality displays, Phrase-DD is a clear winner.

Using similar reasoning, Over-Audio and Phrase-DS appear to be the best choices if audio channel can be utilized. However, Over-Audio needs a microphone on one device (devices such as laptops might not always be equipped with a microphone) and a speaker on the other. Phrase-DS can also be used in scenarios where one device has a speaker. Phrase-SS is also error-free, however, it relatively slow compared to Phrase-DS, and thus there is no good motivation for Phrase-SS to be used over Phrase-DD or Phrase-DS.

Although the methods BEDA-Beep, BEDA-Blink and Beep-Blink have lower hardware requirements and work on devices that just have the most basic interfaces like a beeper, LED or a button, they take too long to complete. They usually require more than one trial to achieve a successful pairing (as we show in Figure 4) and Beep-Blink also yields too high fatal error rate. Considering that the devices taking part in two-user scenarios have decent quality interfaces, BEDA-Beep,

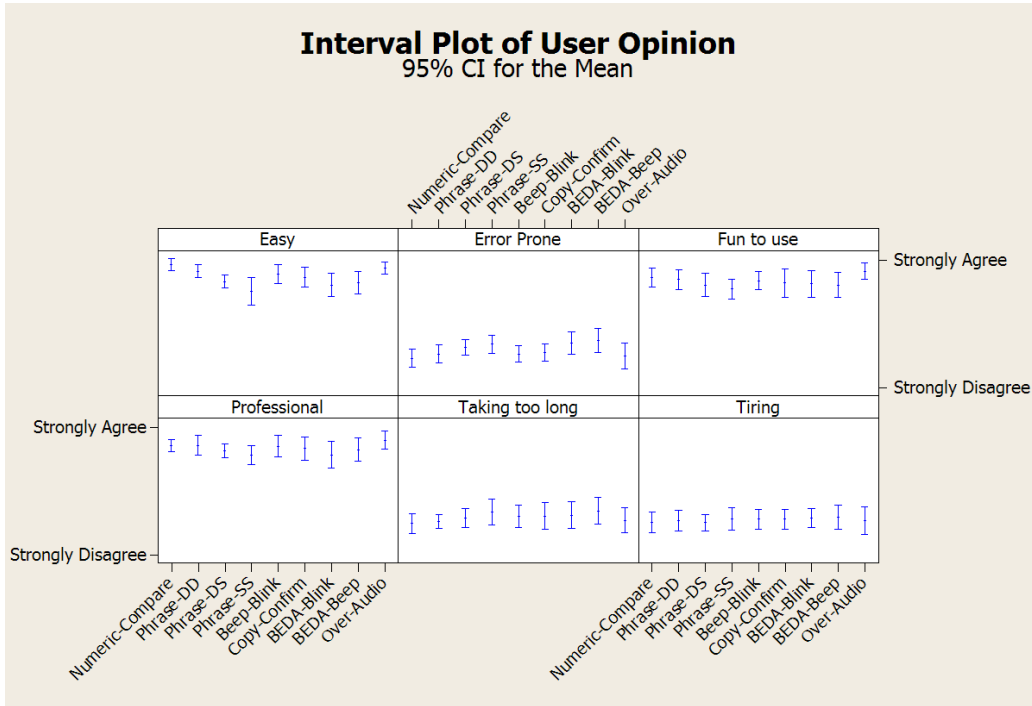

Figure 5: User opinion

BEDA-Blink and Beep-Blink can thus be safely ruled out, in favor of Phrase-DD or Phrase-DS.

## 5.2 Interpreting User Ratings

We now turn our attention to the graph in Figure 5, which summarizes the user opinions collected via the post-test questionnaire. Users are asked to rate six different statements for each method on a 6 point Likert scale lying between ratings "Strongly Disagree" and "Strongly Agree." The rated statements for each method were based on the criteria: {*Easy, Professional, Fun to Use, Tiring, Taking Too Long, Error Prone*}.

As can be seen from Figure 5, there is some observable, and statistically significant, correlation among the user ratings. Methods that are rated as easy are also rated as fun to use and professional. Moreover, methods that are rated as "not easy," were perceived as error prone, tiring and taking too long to complete. The cross correlation of user ratings are given in table 1.

Not surprisingly, Numeric-Compare is ranked among the easiest methods concordant to its fast timing and it being the most familiar method to users as it has already been deployed in many personal devices.

As expected, Phrase-DD and Over-Audio also received very high ratings and are ranked among the easiest, most fun to use and professional methods. All three of Numeric-Compare, Phrase-DD and Over-Audio are among the user favorites.

Contrary to their poor timing and/or high error rates, Beep-Blink and Copy-Confirm are ranked surprisingly positively by our users. Both methods were perceived as easy and error-resistant. Considering Copy-Confirm had a very high fatal error rate (as discussed previously) and rated as one of the least error-prone methods by the subjects, we can conclude that the participants that committed a fatal error were clearly not aware of it. Also judging from the high error rates of both Copy-Confirm and Beep-Blink, we can say that users' perception of security may be far from reality. This contradiction can also be easily observed in methods Phrase-DS and Phrase-SS, although in the other direction. These two methods are ranked among the most error prone of all tested methods but they yield 0% error-rate in our tests.

BEDA-Beep, BEDA-Blink, Phrase-DS and Phrase-SS are the methods considered relatively hard, error prone, taking long time to complete, less professional and less fun to use. The relatively lower user ratings

|             | Easy   | Tiring | Professional | Long   | Fun    |
|-------------|--------|--------|--------------|--------|--------|
| Tiring      | -0.343 |        |              |        |        |
| Professional| 0.693  | -0.266 |              |        |        |
| Long        | -0.445 | 0.817  | -0.293       |        |        |
| Fun         | 0.666  | -0.318 | 0.737        | -0.378 |        |
| Error-Prone | -0.425 | 0.722  | -0.358       | 0.749  | -0.361 |

$p \ll 0.01$, for all correlations

Table 1: Cross-Correlation of Polled Measures

for BEDA-Beep and BEDA-Blink agree with the long completion timings and high error rates we observed in our tests. However, the user perception was deceptive about Phrase-DS and Phrase-SS, especially in terms of how error prone they are.

### 5.3 Final Inferences and Recommendations

Some conclusions that can be drawn by combining the quantitative and qualitative data we collected are as follows:

- Comparison-based pairing methods utilizing visual channel are preferred by our users. Among those, we recommend Phrase-DD over Numeric-Compare as the former yields lower (in fact, no) errors. Since displays are ubiquitous on personal devices participating in two-user scenarios, Phrase-DD is a clear winner in terms of speed, robustness and user preference, as well as universal deployability.

- Among the methods utilizing audio channel, Over-Audio was the user favorite. Phrase-DS also performed well in our tests by yielding low completion timing and no errors. Phrase-SS was also error-free, however, relatively slow. We believe that Phrase-DS is a better choice compared to Over-Audio. This is because Over-Audio needs a microphone on one device (devices such as laptops might not always be equipped with a microphone) and a speaker on the other. Another drawback with Over-Audio is that it is hard for the users to accurately determine the actual source of automated audio, and thus an automated audio channel might be insecure in the presence of a close-by attacker with an interfering audio device [25].[6]

- Beep-Blink and Copy-Confirm produced high error rates in our tests and thus we do not recommend using them.

- BEDA-Beep and BEDA-Blink demonstrated poor usability in our tests. They usually take more than one trial for successful completion and take too long to complete. User ratings for these methods are also relatively low and therefore we do not recommend using them.

- In general, we recommend the use of Phrase-DD when possible and it to be complemented with Phrase-DS methods to be able to accommodate a wider variety of devices with different interface capabilities.

## 6 Conclusions and Future Work

We presented the first experimental evaluation of prominent device pairing methods that can be used in two-user scenarios. First and foremost, our survey results confirmed our belief that a majority of users are considerate about the privacy of their personal devices and they would not be willing to hand-in their devices to other users, even temporarily to perform the pairing process. This means that a two-user pairing method can not be simply reduced to single-user pairing.

The results of our usability study show that one simple method, Phrase-DD, is quite attractive overall, being both fast and error-tolerant as well as user-friendly. It naturally appeals to two-user settings where devices have appropriate quality and size displays. Slightly slower method, Phrase-DS, can seamlessly inter-operate with Phrase-DD, for wider deployment and for scenarios where one device has a speaker. We also observed that, in general, the subjects often decided the outcome of pairing based on mutual agreement, which, we believe, may have helped to reduce errors in most of our comparison-based methods.

---

[6]To detect the presence of a malicious interfering audio device, it would be necessary for the users to perform the manual comparison of SAS strings on both devices, e.g., by using Phrase-DD or Phrase-DS[25]. This undermines the advantage of using an automated audio channel at all.

We believe that our work is an important and timely first step in exploring real-world usability of secure device pairing methods for emerging two-user scenarios. Our work helps affirmatively answer the following important question: *what pairing method(s) should one deploy in practice for two-user scenarios?* Without such a study, it would have been hard to answer this question, based on intuition or prior test results on single-user pairing scenarios. As our results show, what works well for ordinary users is often quite different from what is imagined by researchers. Moreover, usable security is a tricky subject where the user perception may be far from the reality, as our results indicate.

In our future work, we plan on rigorously studying the promising methods resulting from our current study. Our plans for immediate future work include :

- Experiments with more diverse pool of participants, especially those who are not young and technology-savvy, like our current participants.

- Experiments involving different (more diverse) devices (such as laptops, PDAs).

- Experiments in more realistic usage scenarios, outside of a lab-controlled environment.